\documentclass[aps,preprint,preprintnumbers,amsmath,amssymb,epsfx]{revtex4}
\usepackage{graphicx}
\usepackage{dcolumn}
\usepackage{bm}
\usepackage[bookmarks=false]{hyperref}

\def\csf#1#2#3{Chaos, Solitons and Fractals,  {\bf #1}, (#2) #3}
\def\biomj#1#2#3{Biom. J. {\bf #1}, (#2) #3}
\def\ptp#1#2#3{{Prog. Theor. Phys.} {\bf #1}, (#2) #3}

\def\physd#1#2#3{{ Physica D}  #1 (#2) #3}
\def\prl#1#2#3{{ Phys. Rev. Lett.} #1 (#2) #3}
\def\pla#1#2#3{{ Phys. Lett. A} #1 (#2) #3}
\def\pra#1#2#3{{ Phys. Rev. A} #1 (#2) #3}

\def\pre#1#2#3{{ Phys. Rev. E} #1 (#2) #3}

\def\eqn{\end{equation}\noindent}
\def\eqnr{\end{eqnarray}\noindent}
\def\beqr{\begin{eqnarray}}
 \def\beq{\begin{equation}}
\def\chaos#1#2#3{CHAOS,  #1 (#2) #3}

\begin{document}
\title{Experimental realization of mixed-synchronization in counter-rotating coupled oscillators}
\author{
Amit Sharma and Manish Dev Shrimali}
\affiliation{
The LNM Institute of Information Technology, Jaipur 302 031, India \\
}

\begin{abstract}

Recently, a novel {\it \textbf{mixed--synchronization}} phenomenon is observed in counter--rotating nonlinear coupled oscillators \cite{r0}. In mixed--synchronization state: some variables are synchronized in--phase, while others are out--of--phase. We have experimentally verified the occurrence of mixed--synchronization states in coupled counter--rotating chaotic piecewise $R\ddot{o}ssler$ oscillator. Analytical discussion on approximate stability analysis and numerical confirmation on the experimentally observed behavior is also given.

\end{abstract}
\maketitle

\section{Introduction}

Huygens first describes the anti--phase synchronization in a pair of pendulum clocks \cite{ch}. Later, the idea of synchronization of two identical chaotic system was introduced by Pecora and Carroll~\cite{r1}. Synchronization of chaotic systems has attracted much attention due to its potential application in secure communication, chemical and biological system, information science, and so on \cite{r1a}. Many different synchronization states have been studied in literature, namely complete or identical synchronization (CS) \cite{r1,r3,r4}, in--phase (PS) \cite{r6,r7}, anti--phase \cite{r8}, lag synchronization (LS)~\cite{r9}, generalized synchronization (GS) \cite{r10,r11}, intermittent lag synchronization(ILS) \cite{r12,r13}, and anti-synchronization (AS) \cite{r14,r15,r16} in which one of the dynamical variable is synchronized then rest of variable follow the same. All these type of synchronization can be achieved with various type of interactions e.g. mismatch oscillators \cite{r17}, conjugate \cite{r19, r19a}, delay \cite{r18}, and nonlinear \cite{r20, r20a}, indirect \cite{raa, amit}.

If the directions of rotation of two oscillators are same, the system is co--rotating, while system of oscillators rotating in opposite direction is called counter--rotating. Coupled co--rotating nonlinear oscillators have been extensively studied from both theoretical and experimental point of view \cite{r1a}.
Recently, a {\it mixed--synchronization} phenomenon was observed in coupled counter--rotating nonlinear oscillators \cite{r0}, similar phenomena was engineered using a general formulation of coupling function in co--rotating coupled oscillators \cite{r26, r26a}.
In mixed--synchronization state, some variables are synchronized to in--phase state while other variables are out--of--phase. The mixed--synchronization phenomenon is also studied in the case of extended systems \cite{r0}.

In this Letter, we present the experimental observation of the mixed--synchronization in two diffusive coupled counter--rotating chaotic piecewise $R\ddot{o}ssler$ oscillators. The analytical discussion on approximate stability analysis and numerical simulations are in close agreement with experimental results. The critical value of coupling strength, where counter--rotating coupled chaotic oscillators are synchronized, is larger in experiments as compared to numerical simulations because of parameter mismatch in circuit implementation.

The Letter is organized as follows: In the section II we numerically study the mixed--synchronization phenomenon in the coupled chaotic  oscillators for piecewise $R\ddot{o}ssler$ system. The linear stability analysis and numerical results are presented. The experimental setup and the results of coupled counter--rotating chaotic oscillators are discussed in section III. Concluding remarks and discussion about the {\it mixed--synchronization} in coupled counter--rotating chaotic oscillators are given in section IV.

\section{The model system}

Here, we illustrate the mixed--synchronization phenomena in two diffusive coupled piecewise $R\ddot{o}ssler$ \cite {r23} systems given by following equations

\begin{eqnarray}
\dot{x_i} &=& -\gamma_ix_i-\alpha_iy_i -z_i+ \epsilon(x_j - x_i) = f(x_i, y_i, \omega_i) + \epsilon(x_j - x_i) \nonumber \\
\dot{y_i} &=& \beta_ix_i+a_iy_i = g(x_i, y_i, \omega_i) \nonumber\\
\dot{z_i} &=& h(x_i)-z_i
\label{eq:eq1}
\end{eqnarray}

Here, in functions $f$ and $g$, $\omega_i$'s represents the internal frequency of two oscillators with opposite sign and it depends on the parameters $\alpha_i$ and $\beta_i$. The function $h(x)=0$ if $x \le 3$, or $h(x)=\mu(x-3)$ if $x > 3$. The rotation of Piecewise $R\ddot{o}ssler$ system (in $x-y$ plane) can be changed by changing the sign of $\alpha_i$ and $\beta_i$. The first system has counter clockwise rotation while second has clockwise rotation.
The parameters values are: $\alpha_1=0.5$, $\alpha_2=-0.5$, $\beta_1=1$, $\beta_2=-1$, $\gamma_i=0.05$, $a_i=0.113$, and $\mu_i=15$. The coupling parameter is $\epsilon$. For identical oscillators, $a_1=a_2$.

The fixed points of the piecewise $R\ddot{o}ssler$ oscillators are $(x^{*}=\frac{3\mu_i}{\kappa_i},y^{*}=\frac{-3\mu_i\beta_i}{a_i\kappa_i}, z^{*}=\frac{-3\mu_i}{\kappa_i}(-\gamma_i+\frac{\alpha_i \beta_i}{a_i}))$, where $\kappa_i=-\gamma_i+\frac{\alpha_i\beta_i}{a_i}+\beta_i$ depends on the sign of the $\alpha_i$ and $\beta_i$.
The change in the dynamical behavior arises from the coupling between two identical piecewise $R\ddot{o}ssler$ oscillators.

\subsection{Numerical Results}

\begin{figure}
\includegraphics[width=0.8\textwidth]{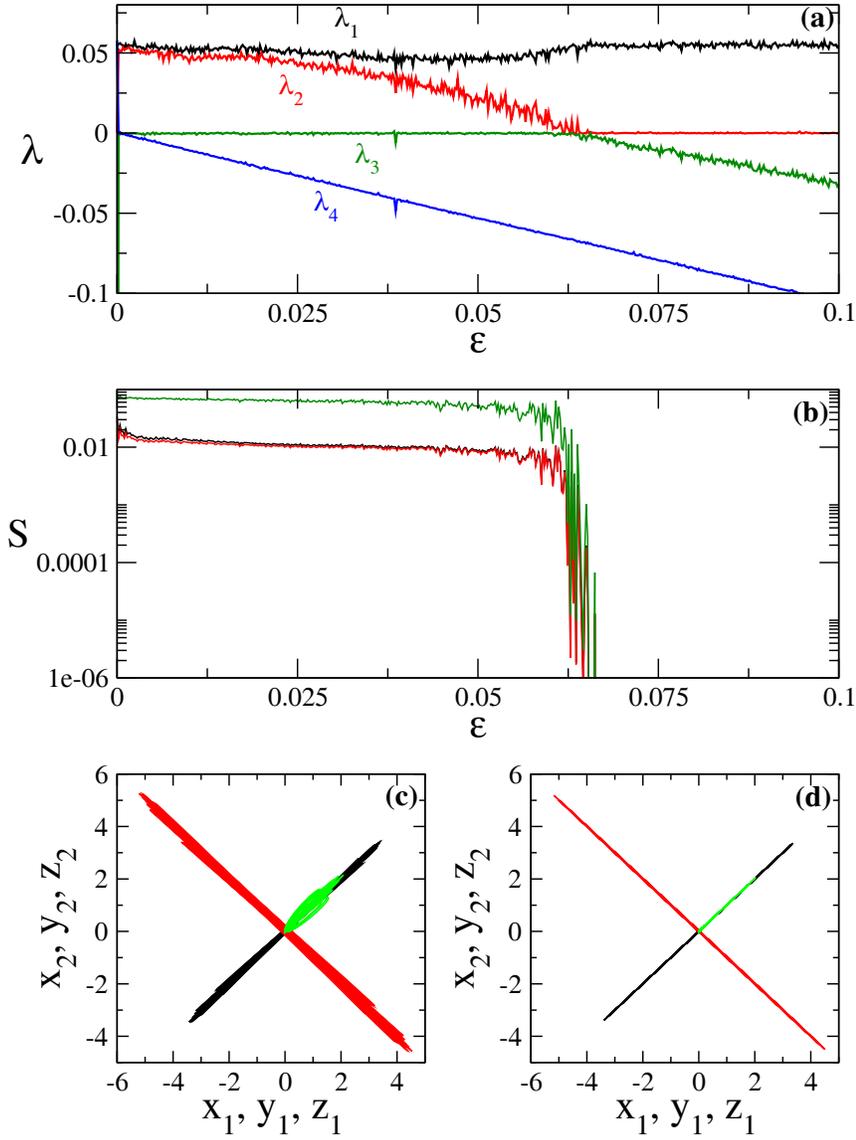}
\vskip 0.5cm
\caption{ (Color online) (a) The largest four Lyapunov exponents of identical coupled counter--rotating piecewise $R\ddot{o}ssler$ oscillators. (b) the similarity functions for $x, y$ and $z$ variables of the coupled oscillators. (c) and (d) the phase relationship between the variables $x, y$ and $z$ at $\varepsilon=0.06$ and $\varepsilon=0.1$ respectively.
The similarity function and dynamics for variables $x, y$, and $z$ are marked by black, red, and green color respectively.}
\vskip1cm
\label{fig:fig1}
\end{figure}

We numerically study the mixed--synchronization of two coupled counter--rotating piecewise $R\ddot{o}ssler$ oscillators. At the very small coupling strength the two oscillators are uncorrelated. As the coupling strength increases, the phase synchronization set in when forth largest Lyapunov exponent becomes negative and complete synchronization occurs when third largest Lyapunov exponent becomes negative as shown in Fig.~\ref{fig:fig1}(a).

To quantify synchronization, we use the following similarity function defined with respect to dynamical variables, $x$ and $y$, of the chaotic oscillator \cite{r9}
\beq
S(x) = \sqrt \frac{<[x_2(t)-x_1(t)]^2>}{[<x_1^2(t)><x_2^2(t)>]^{1/2}}
\label{eq:eq10}
\eqn

\beq
S(y) = \sqrt \frac{<[y_2(t)+y_1(t)]^2>}{[<y_1^2(t)><y_2^2(t)>]^{1/2}}
\label{eq:eq11}
\eqn

Synchronization (complete and anti) is characterized by $S(.)=0$ for $x$ and $y$ variables respectively.
The variables $x_1$ and $x_2$ are in--phase while $y_1$ and $y_2$ are out--of--phase. The two variables $x$ and $y$ shows complete in--phase and out--of--phase synchronization respectively for coupling strength $\epsilon > \epsilon_c$, where $\epsilon_c \sim 0.067$. The $z$ variable of the system also goes to complete synchronization state, where $S(z)$ is defined similar to $S(x)$.
The in--phase synchronization in $x_1$ and $x_2$ while out--of--phase in $y_1$ and $y_2$ is shown in Figure~\ref{fig:fig1}(c). The complete synchronization in $x_1$ and $x_2$ with zero relative phase while out--of--phase state of $y_1$ and $y_2$ with phase difference of $\pi$ is shown in Figure~\ref{fig:fig1}(d).

\subsection{Linear Stability Analysis}




We analyze the stability of the mixed--synchronized state of two counter--rotating coupled chaotic systems  given by Eq.~(\ref{eq:eq1}) in $x-y$ plane. The method of approximate linear stability analysis is adopted for synchronization
criteria \cite{raa}. If $\xi$ and  $\eta$ represent the deviation of coordinates $x$ and $y$ respectively from the synchronization state, their dynamic is governed by the linearized equation as

\begin{eqnarray}
\dot{\xi_i} &=& f'(x_i,y_i,\omega_i) + \epsilon(\xi_j - \xi_i), \nonumber\\
\dot{\eta_i} &=& g'(x_i,y_i,\omega_i)
\label{eq:eq2}
\end{eqnarray}

Where the $f$ and $g$ are functions in terms of coordinate and parameter. $\omega_i$, i=1,2 represent the frequency of the oscillators. The criteria for the stability is that synchronization state corresponding to fixed point will be stable if all eigen values of the Eqs.~(\ref{eq:eq2}) are negative.

Dynamics of the deviation from the synchronization state is governed by the linearized equation of Eqs~(\ref{eq:eq1}).

\begin{eqnarray}
\dot{\xi_1} &=& (- \gamma \xi_1 -\alpha_1 \eta_1)f'(x_1,y_1) +\epsilon(\xi_2 - \xi_1), \nonumber \\
\dot{\eta_1} &=& (\beta_1\xi_1 + a \eta_1)g'(x_1,y_1), \nonumber\\
\dot{\xi_2} &=& (- \gamma \xi_2 -\alpha_2 \eta_2)f'(x_2,y_2) +\epsilon(\xi_1 - \xi_2), \nonumber \\
\dot{\eta_2} &=& (\beta_2\xi_2 + a \eta_2)g'(x_2,y_2)
\label{eq:eq3}
\end{eqnarray}

Where $\gamma, \alpha, \beta$, and $a$ are parameters. For the Perfect synchronization in counter rotating coupled system , i.e. $x_1 = x_2$ (complete) and $y_1 = -y_2$ (Anti-synchronization), we can define

\begin{eqnarray}
\mu_1 &=& \xi_1 - \xi_2, \nonumber\\
\mu_2 &=& \eta_1 + \eta_2
\label{eq:eq4}
\end{eqnarray}

Then Eqs~(\ref{eq:eq3}) can be written as

\begin{eqnarray}
\dot{\mu_1} &=& -\gamma f'(x_1,y_1)\mu_1-(\alpha_1\eta_1-\alpha_2\eta_2)f'(x_1,y_1)-2\epsilon\mu_1,\nonumber\\
\dot{\mu_2} &=& (\beta_1\xi_1+\beta_2\xi_2)g'(x_1,y_1) + a g'(x_1,y_1)\mu_2
\label{eq:eq5}
\end{eqnarray}

If we assume that the time average values of Jacobian matrix elements $f'(x_i,y_j)$ and $g'(x_i,y_i)$, where i=1,2 are approximately the same and can be replaced by an effective constant value $\lambda_1$ and $\lambda_2$.

In the case of counter rotating $R\ddot{o}ssler$ systems, frequency of the coupled systems are of opposite sign: $\alpha_1 = -\alpha_2 $ and $\beta_1 = -\beta_2$. Then

\begin{eqnarray}
\dot{\mu_1} &=& -(\gamma\lambda_1+2\epsilon)\mu_1-\alpha_1\lambda_1\mu_2 ,\nonumber\\
\dot{\mu_2} &=& \beta_1\lambda_2\mu_1+a \lambda_2\mu_2
\label{eq:eq6}
\end{eqnarray}

Eliminating $\mu_2$  from above equations, we get

\begin{eqnarray}
\ddot{\mu_1} &=& (a\lambda_2-(\gamma\lambda_1+2\epsilon))\dot{\mu_1}-(\alpha_1\beta_1\lambda_1\lambda_2-a\lambda_2(\gamma\lambda_1+2\epsilon))\mu_1
\label{eq:eq7}
\end{eqnarray}

Solution of the equation $\mu_1 = Ae^{mt}$, we get

\begin{eqnarray}
m &=& \frac{(a\lambda_2-(\gamma\lambda_1+2\epsilon))\pm\sqrt{(a\lambda_2-(\gamma\lambda_1+2\epsilon)))^2 - 4(\alpha_1\beta_1\lambda_1\lambda_2-a\lambda_2(\gamma\lambda_1+2\epsilon))}}{2}
\label{eq:eq8}
\end{eqnarray}

The synchronization state define by $\mu_1 = \xi_1-\xi_2=0$ and $\mu_2 = \eta_1+\eta_2=0$, is stable if Re[\emph{m}] is negative for both the solutions.

\begin{itemize}
  \item If $(a\lambda_2-\gamma\lambda_1-2\epsilon))^2 < 4(\alpha_1\beta_1\lambda_1\lambda_2-a\lambda_2(\gamma\lambda_1+2\epsilon))$, \emph{m} is complex and the stability condition becomes $(\gamma\lambda_1+2\epsilon) > a\lambda_2$.
  \item If $(a\lambda_2-\gamma\lambda_1-2\epsilon))^2 > 4(\alpha_1\beta_1\lambda_1\lambda_2-a\lambda_2(\gamma\lambda_1+2\epsilon))$, \emph{m} real and the stability condition becomes $(\gamma\lambda_1+2\epsilon) > a\lambda_2$.
\end{itemize}

The transition to stable synchronization is given by the threshold values of the parameters satisfying the condition

\beq
\epsilon_c=\frac{1}{2}(a \lambda_2 - \gamma \lambda_1)
\label{eq:eq9}
\eqn

\begin{figure}
\includegraphics[width=0.8\textwidth]{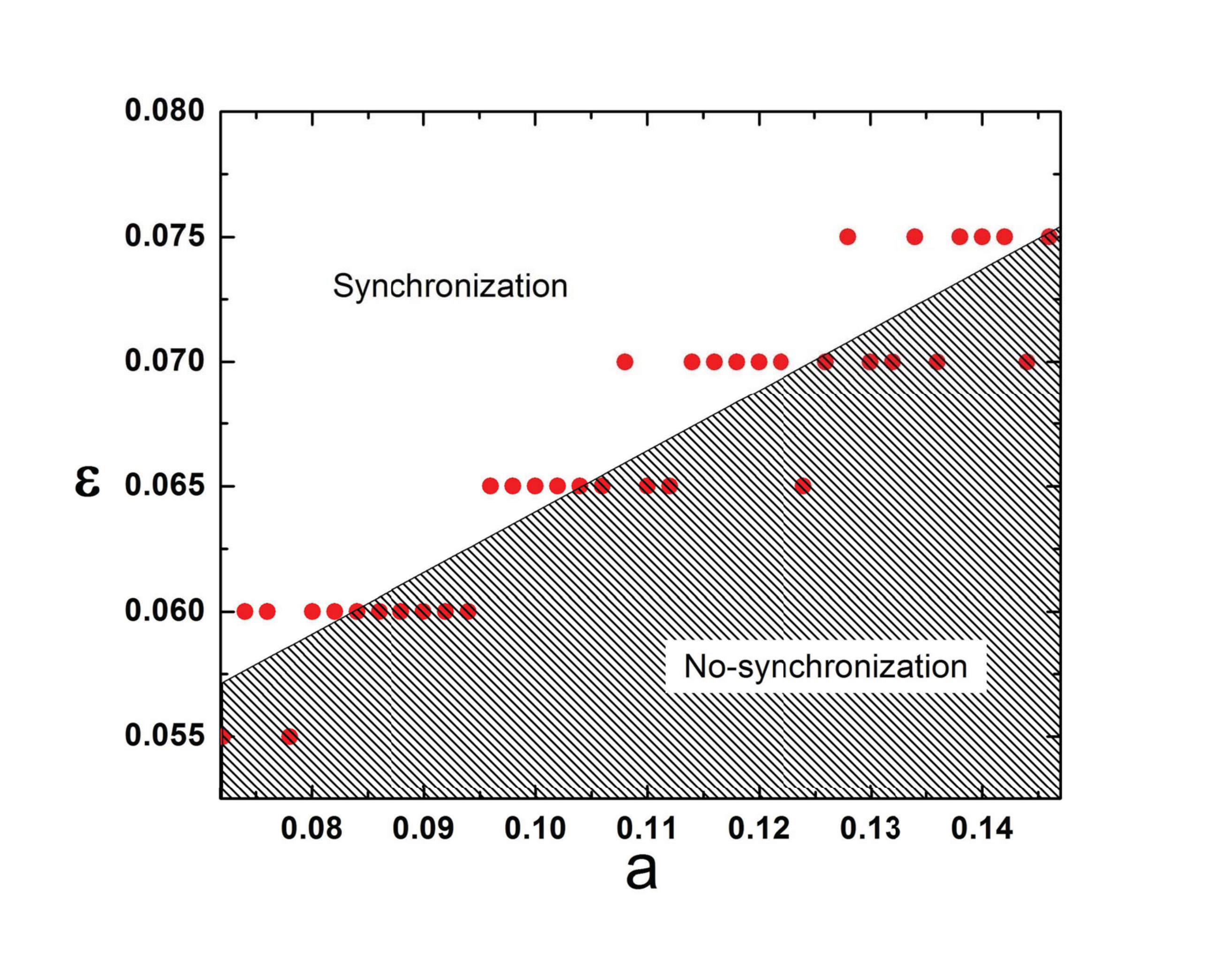}
\vskip 1.5cm
\caption{(Color online)
Transition from unsynchronized to mixed--synchronized region is shown in the parameter plane $(a, \epsilon)$ for coupled piecewise $R\ddot{o}ssler$ oscillators.
}
\label{fig:fig2}
\end{figure}

Figure~\ref{fig:fig2} shows the transition from the unsynchronized to mixed--synchronization state in the $\epsilon - a$ space. A linear relations is clearly seen and the solid line is drawn with the effective $\lambda_1=-1.45$ and $\lambda_2=0.55$, thus validating the transition criterion of Eq.~(\ref{eq:eq9}) obtained from the stability theory. The condition given by Eq.~(\ref{eq:eq9}) is necessary but not sufficient for synchronization.

\section{Experimental Setup and Results}

Experiments are conducted using a pair of electronic oscillators whose dynamics mimic that of the chaotic $R\ddot{o}ssler$ oscillator \cite{r23}. One of the oscillators rotate clockwise while another anti--clockwise. The two oscillators are approximately identical since in reality it is not possible to ensure that parameters are exactly equal. Further, unlike the piecewise $R\ddot{o}ssler$ system (Eq.~(\ref{eq:eq1})) discussed above, the coupling is asymmetric and frequencies of the oscillators are not equal in experiment. Hence, we observe lag and phase synchronization in coupled piecewise $R\ddot{o}ssler$ oscillators as coupling is increased.

The Piecewise $R\ddot{o}ssler$ oscillator circuit shows the dynamics of rotation in counter clockwise.
We have to connect two inverting amplifier $U_9$ and $U_{10}$ (as shown in Fig.~3) for changing the direction of the rotation in circuit. Both piecewise $R\ddot{o}ssler$ oscillators are consisting of the passive components like resistance $R_{1-42}$, capacitors $C_{1-6}$, diodes $D_{1-2}$ and operational amplifier uA741 $U_{1-16}$. We use simple linear scheme for the coupling between $x$ variables of the two piecewise $R\ddot{o}ssler$ oscillators. The OPAMP $U_{6,7,15,16}$ in circuit are used for linear coupling scheme. $RF_1$ and $RF_2$ are the variable resistors characterizing the coupling parameter.
The electronic components in circuits are carefully chosen and values are mentioned in the diagram (Fig.~3). The typical oscillating frequencies of the circuits are in the audio frequency range. Both oscillators are operated by a low-ripple and low noise power supply. The output voltages form both oscillators are monitored using digital oscilloscope 100MHz 2 channel (Agilent DSO1012A) with maximum sampling rate of 2 GSa/s.

Transitions from asynchronous chaos to lag synchronization and then to in--phase synchronization is observed at the critical values of variable resistance, $R_{c1}$ and $R_{c2}$, respectively. The lag synchronization occurs in the interval [$R_{c1}$, $R_{c2}$], where $R_{c1}$ = 0.5k$\Omega$ and $R_{c2}$ = 9k$\Omega$.
It has been observed in experiments that the variables of one of the oscillator tends to fellow the variables of the another oscillator in some range of the coupling strength \cite{A21}. Here, it is due to the parameter mismatch in the coupled oscillators.
The Similarity function of $x$ and $y$ variables (Eq.~(\ref{eq:eq10}) and (\ref{eq:eq11})) of the coupled piecewise $R\ddot{o}ssler$ oscillator with variable resistance $R$ is shown in Fig.~4. At $RF_1$ = $RF_2$ = 1.4k$\Omega$ the output voltage of $x_1$ and $x_2$ shows in--phase dynamics while $y_1$ and $y_2$ are out--of--phase. Phase relationship of $x$ and $y$ variables with lag synchronization are shown in Fig.~5(c-d). Further increase of the coupling strength shows the transition from lag to phase synchronization. Phase relationship at $RF_1$ = $RF_2$ = 11.2k$\Omega$ is shown in Fig.~5(e-f).

\section{Conclusion}

We presented the experimental evidence of mixed--synchronization in the piecewise $R\ddot{o}ssler$ oscillators circuit via diffusive type of coupling under the parameter mismatch. The experimental results are in close agreement with the numerical results. The critical value of coupling strength for onset of mixed--synchronization is calculated using approximate linear stability analysis.
We have also studied the sprott circuit \cite{r24} and obtained similar results for mixed-synchronization. The natural emergence of novel {\it mixed--synchronization} phenomenon in chaotic as well as limit cycle counter--rotating coupled oscillators has possible applications in secure communication and chaos based computing.

\section*{Acknowledgments}

We thank Awadhesh Prasad and Syamal Kumar Dana for useful discussion and critical comments on the manuscript. We would like to acknowledge the financial support from DST, India and LNMIIT, Jaipur.

\newpage

{\bf Figure Captions}

\vskip 1cm

\noindent
{\bf Figure 1:} (Color online) (a) The largest four Lyapunov exponents of identical coupled counter--rotating piecewise $R\ddot{o}ssler$ oscillators. (b) the similarity functions for $x, y$ and $z$ variables of the coupled oscillators. (c) and (d) the phase relationship between the variables $x, y$ and $z$ at $\varepsilon=0.06$ and $\varepsilon=0.1$ respectively.
The similarity function and dynamics for variables $x, y$, and $z$ are marked by black, red, and green color respectively. \\

\noindent
{\bf Figure 2:} (Color online) Transition from unsynchronized to mixed--synchronized region is shown in the parameter plane $(a, \epsilon)$ for coupled piecewise $R\ddot{o}ssler$ oscillators. \\

\noindent
{\bf Figure 3:} (Color online) Schematic diagram of two bidirectional coupled (counter clockwise and clockwise) piecewise chaotic $R\ddot{o}ssler$ oscillator.
Variable resistors are used to change the coupling. The OPAMP are type of uA741. All resistors are metal  film type with tolerance $1 \%$ and capacitors are polyester type with tolerance $5 \%$. The circuit is run by $\pm12V$ source. \\

\noindent
{\bf Figure 4:} (Color online) Similarity function $S$ for $x$ (circle) and $y$ (triangle) variables of the  experimental system of two coupled counter rotating piecewise $R\ddot{o}ssler$ oscillators with variable resistance $RF_1 = RF_2 = R$. \\

\noindent
{\bf Figure 5:} (Color online) Dynamic of the piecewise $R\ddot{o}ssler$ oscillator in (a) counter clockwise rotation (b) clockwise rotation. The phase relationship of $x$ and $y$ variables respectively at $RF_1 = RF_2 = 1.4k \Omega$ for lag--synchronization in (c) and (d). mixed--synchronization at $RF_1 = RF_2 = 11.2k \Omega$ in (e) and (f).\\

\newpage

\begin{figure}
\includegraphics[width=0.8\textwidth]{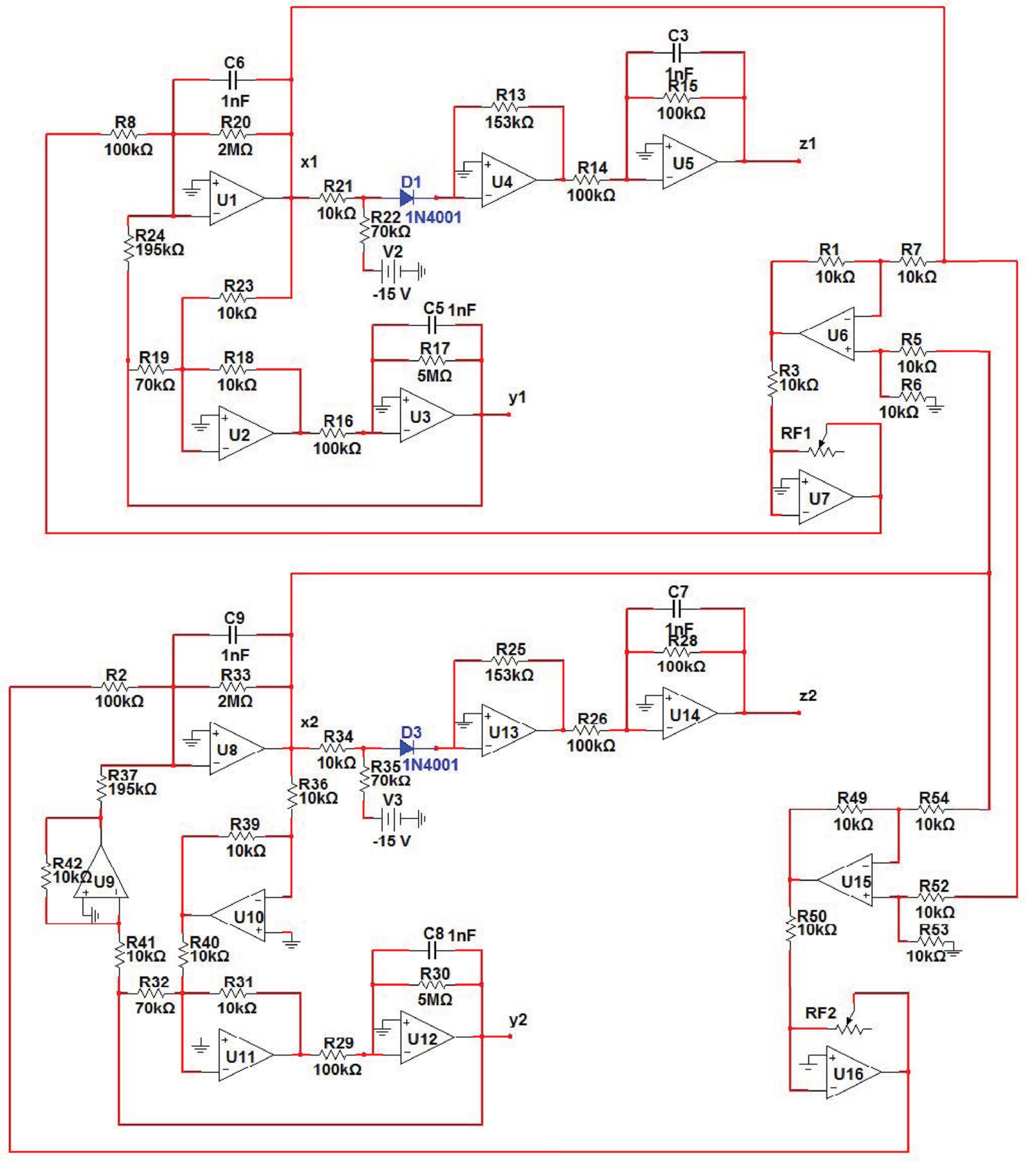}
\vskip 1.5cm
\caption{(Color online)
Schematic diagram of two bidirectional coupled (counter clockwise and clockwise) piecewise chaotic $R\ddot{o}ssler$ oscillator.
Variable resistors are used to change the coupling. The OPAMP are type of uA741. All resistors are metal  film type with tolerance $1 \%$ and capacitors are polyester type with tolerance $5 \%$. The circuit is run by $\pm12V$ source.
}
\label{fig:fig3}
\end{figure}

\begin{figure}
\includegraphics[width=0.8\textwidth]{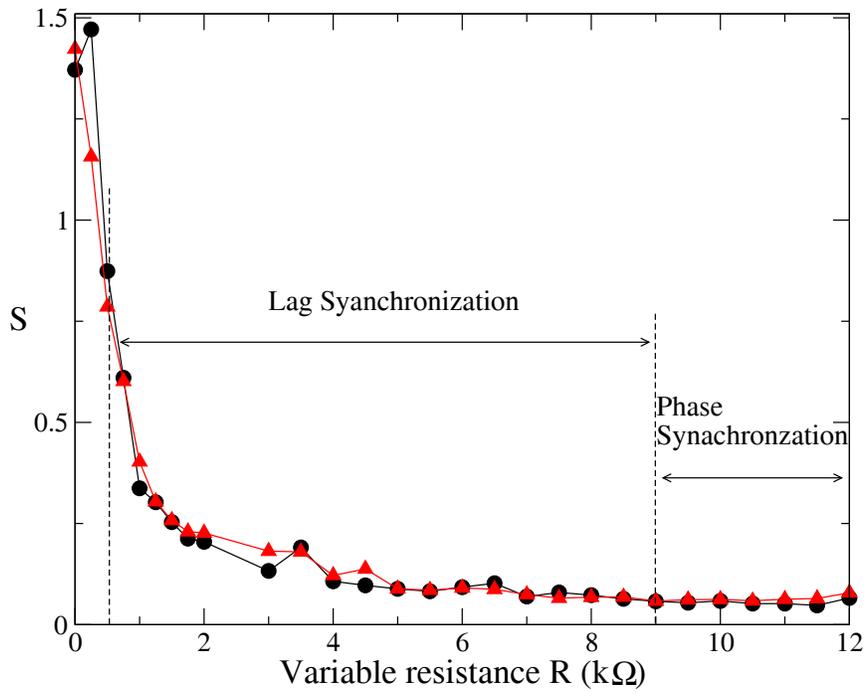}
\vskip 3.cm
\caption{(Color online)
Similarity function $S$ for $x$ (circle) and $y$ (triangle) variables of the  experimental system of two coupled counter rotating piecewise $R\ddot{o}ssler$ oscillators with variable resistance $RF_1 = RF_2 = R$.
}
\label{fig:fig4}
\end{figure}

\begin{figure}
\includegraphics[width=0.8\textwidth]{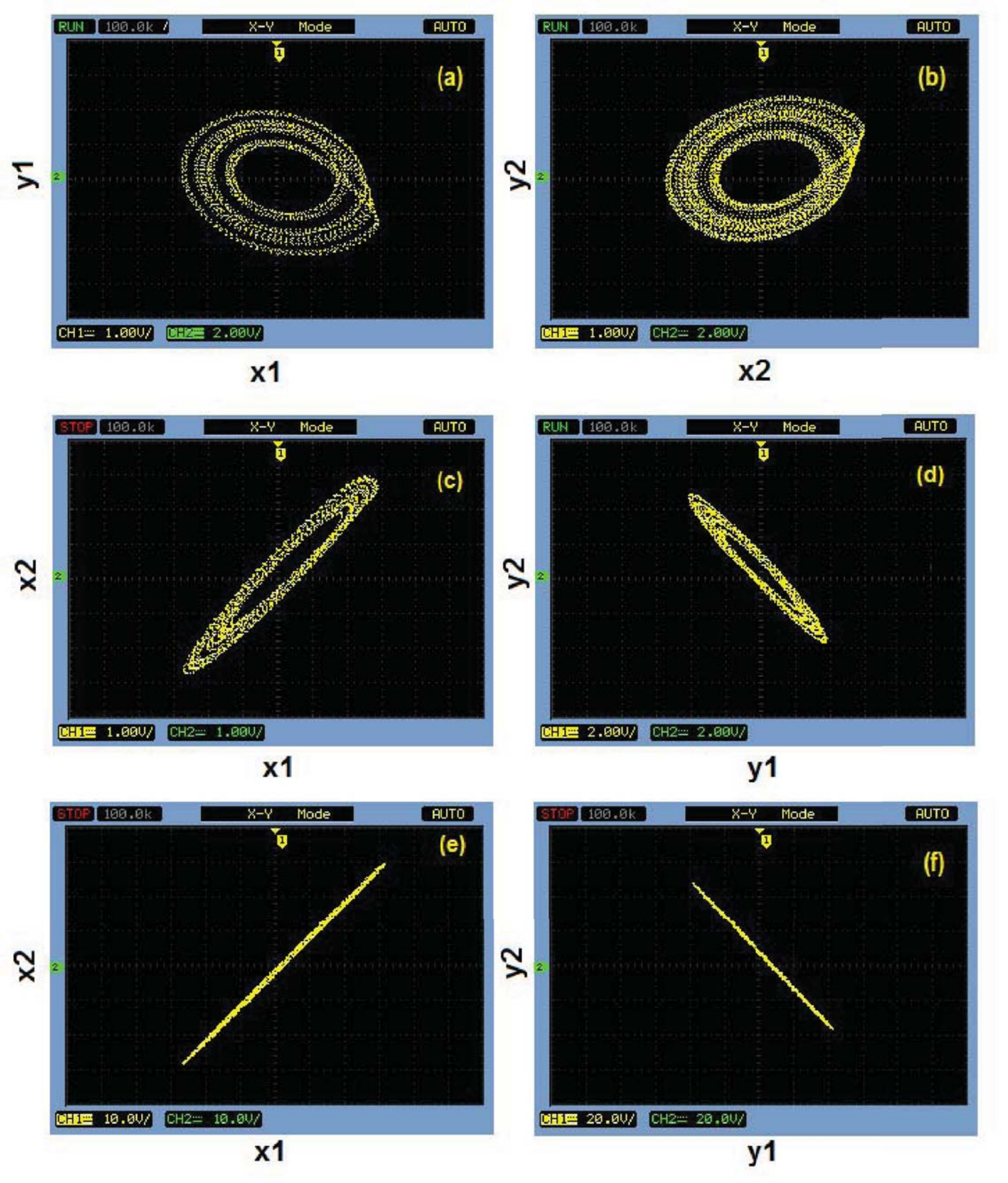}
\vskip 3.cm
\caption{(Color online)
Dynamic of the piecewise $R\ddot{o}ssler$ oscillator in (a) counter clockwise rotation (b) clockwise rotation. The phase relationship of $x$ and $y$ variables respectively at $RF_1 = RF_2 = 1.4k \Omega$ for lag--synchronization in (c) and (d). mixed--synchronization at $RF_1 = RF_2 = 11.2k \Omega$ in (e) and (f).}
\label{fig:fig5}
\end{figure}

\end{document}